# Electronic transport through ordered and disordered graphene grain boundaries

Péter Vancsó[1,4*], Géza I. Márk[1,4], Philippe Lambin[2], Alexandre Mayer[2], Yong-Sung Kim[3,4], Chanyong Hwang[3,4], and László P. Biró[1,4]

[1]Institute of Technical Physics and Materials Science, Centre for Natural Sciences, H-1525 Budapest, P.O. Box 49, Hungary (http://www.nanotechnology.hu/)

[2]Physics Department, University of Namur (FUNDP), 61 Rue de Bruxelles, B-5000 Namur, Belgium

[3]Center for Nano-characterization, Division of Industrial Metrology, Korea Research Institute of Standards and Science, Yuseong, Daejeon 305-340, Republic of Korea

[4]Korean-Hungarian Joint Laboratory for Nanosciences, H-1525 Budapest, P.O. Box 49, Hungary

**Abstract**

The evolution of electronic wave packets (WPs) through grain boundaries (GBs) of various structures in graphene was investigated by the numerical solution of the time-dependent Schrödinger equation. WPs were injected from a simulated STM tip placed above one of the grains. Electronic structure of the GBs was calculated by *ab-initio* and tight-binding methods. Two main factors governing the energy dependence of the transport have been identified: the misorientation angle of the two adjacent graphene grains and the atomic structure of the GB. In case of an ordered GB made of a periodic repetition of pentagon-heptagon pairs, it was found that the transport at high and low energies is mainly determined by the misorientation angle, but the transport around the Fermi energy is correlated with the electronic structure of the GB. A particular line defect with zero misorientation angle [Lahiri et al, Nat Nanotechnol 2010;5:326–9] behaves as a metallic nanowire and shows electron-hole asymmetry for hot electrons or holes. To generate disordered GBs, found experimentally in CVD graphene samples, a Monte-Carlo-like procedure has been developed. Results show a reduced transport for the disordered GBs, primarily attributed to electronic localized states caused by C atoms with only two covalent bonds.

---

[*] Corresponding author. Email-address: vancso.peter@ttk.mta.hu (P. Vancsó)

1. Introduction

Graphene, a single layer of graphite has unique electronic and transport properties[1] resulting from the linear energy dispersion relations of the charge carriers near the Fermi level. Utilizing the electronic and the outstanding thermal and mechanical properties of graphene new perspectives may be opened for future applications[2,3,4]. Nowadays, the chemical vapor deposition (CVD) method[5] on metal surfaces is one of the most promising ways to produce graphene sheets at the scale and quality required for practical applications. Nevertheless, these graphene films are polycrystalline containing several grain boundaries (GBs) between single crystal regions[6]. These defects[7] may substantially affect the remarkable electronic[8,9] and mechanical[10] properties of the perfect graphene lattice. However, they also offer an opportunity to modify the local properties via defect engineering to achieve new functionalities.

Due to the importance of GBs in future applications of graphene, they have been investigated in numerous theoretical calculations[11,12,13]. These works focus on well-ordered periodic structures containing pentagons, heptagons and octagons, which have been observed in HOPG[14]. Calculated ordered GBs usually have a low total energy configuration[10] but according to measurements[15] their electronic structures around the Fermi level can be different from each other.

Electronic transport across the GBs depends on two main parameters: i) the misorientation of the two grains and ii) the detailed atomic- and electronic structure of the GB. Yazyev and Louie[16] have presented a theoretical model based on the zone folding method showing that ordered periodic GBs may open a transport gap around the Fermi level. Unexpectedly, the presence of the transport gap does not depend on the detailed atomic structure of the GB, just on the value of the misorientation angle and the linear periodicity of the GB. The effect of the atomic structure on the transport properties around the Fermi level can be investigated only in structures without a transport gap.

However, recent measurements[9] and computer models[17,18] have shown that GBs on CVD produced graphene may have more amorphous like structures compared with those seen on HOPG. The GBs might contain other polygonal rings[17] like squares and nonagons and their structures are disordered[18]. In addition, they are not perfectly straight lines and are wider than the ordered ones. For a recent review of GBs in CVD graphene see Ref. [19]. The electronic properties[9,20] of the GBs can be also altered by the increasing number of defects in the disordered structure. Vacancies and the two coordinated carbon atoms may cause significant modulations in the local density of states (LDOS) near the Fermi level modifying also the low-energy carriers transport properties. Thus, revealing the effect of these disordered GBs on the transport properties plays an important role in understanding the conductivity of polycrystalline graphene samples.

Wave packet dynamics[21] (WPD) is an effective method to investigate electron propagation in nanostructures like graphene[22,23]. The approach is based on the idea that transport can be viewed as a problem in potential scattering theory[24]. The scattered wave packets (WPs) on a potential barrier are calculated numerically.

In this paper we examine in detail the dynamics of electrons through ordered and disordered GBs in geometries modeling an experimental situation. The electron WPs are injected from a simulated scanning tunneling microscope (STM) tip, which is situated above one of the graphene grains, far from the GB line (Fig. 1), then the time evolution of the injected WP is followed while crossing through the GB.

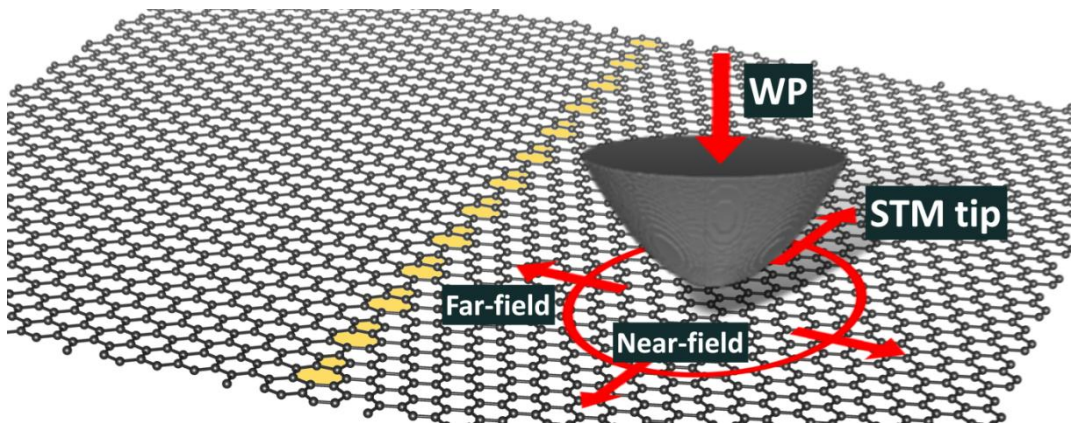

Fig. 1 - Model geometry of the STM tip - graphene system for the case of the pentagon-heptagon GB. STM tip modeled with a rotational hyperboloid of jellium is presented by the -2.7eV equipotential surface of the potential. Red arrows symbolize the incoming and spreading directions of the wave packet. Below the STM tip the red circle denotes the near-field region, where the STM tip has strong influence on the wave functions. See text for details.

A drain potential[25] applied along the edges of the presentation box serves as the other STM electrode. The geometry has the advantage that we do not need to make assumptions about the initial shape of the electron WP on graphene, because it is constructed from the stationary states of the STM tip. Our structural model contains both components affecting the transport phenomenon: the misorientation of the two grains and the local geometry together with the electronic structure of the GB. Compared to other transport methods, WPD is able to handle systems containing a large number of C atoms (more than 10,000) even on a PC, therefore we are not limited to ordered GBs with small periodicity as in the case of *ab-initio* calculations. The WPD method is opening a way for investigating disordered structures observed in the CVD produced graphene samples[9,19,26]. In this paper we performed calculations for two ordered defects: a pentagon-heptagon (5-7) GB, a pentagon-octagon (5-5-8) extended linear defect, where the misorientation angles are 38.2° and zero, respectively, and for a disordered large angle GB with geometry modeled by a Monte-Carlo procedure. According to Yazyev's theory[16] these GBs do not have transport gap around the Fermi level. These simulations allowed us to separate the transport components related to the misorientation (5-7 GB) and to the electronic structure (5-5-8 defect), as well as to identify the signature of disorder on the transport.

## 2. Simulation methods

### 2.1 Construction of the GB structures

The graphene sheet containing a GB has three regions: two perfect graphene lattices rotated with respect to each other and the GB region. In the case of the ordered GBs, the structure near the GB was relaxed by minimizing the Tersoff-Brenner potential[27].

The technique for the construction of disordered GBs works as follows. Two graphene cells are generated on the computer (each cell is a parallelogram sustained by two lattice vectors). By a rotation, an edge of the first cell is made parallel to an edge of the second cell. Subsequently, one cell is translated so as to overlap the other over a region of 1 nm in width (Fig. 2a). This distance will eventually become the width of the grain boundary region. The geometrical border between the grains is defined as the median line of the overlapping zone.

The sites that belong to the overlapping region are emptied. The empty sites of either lattice are then sequentially filled according to the following procedure. Each empty site receives a weight: 0 when its three nearest neighbors are empty, 1 or 2 if respectively one or two nearest-neighbor sites are occupied by C atoms. The weights were chosen so as to maximize the number of three-fold coordinated atoms. An empty site is selected randomly with a probability proportional to the filling weight it has received and it is filled with a C atom. This method assures that the two grains grow from their front edge towards the grain boundary (Fig. 2b). At each step of the procedure, any lattice vacancy (an empty site having its three nearest neighbors occupied by C atoms) is automatically filled.

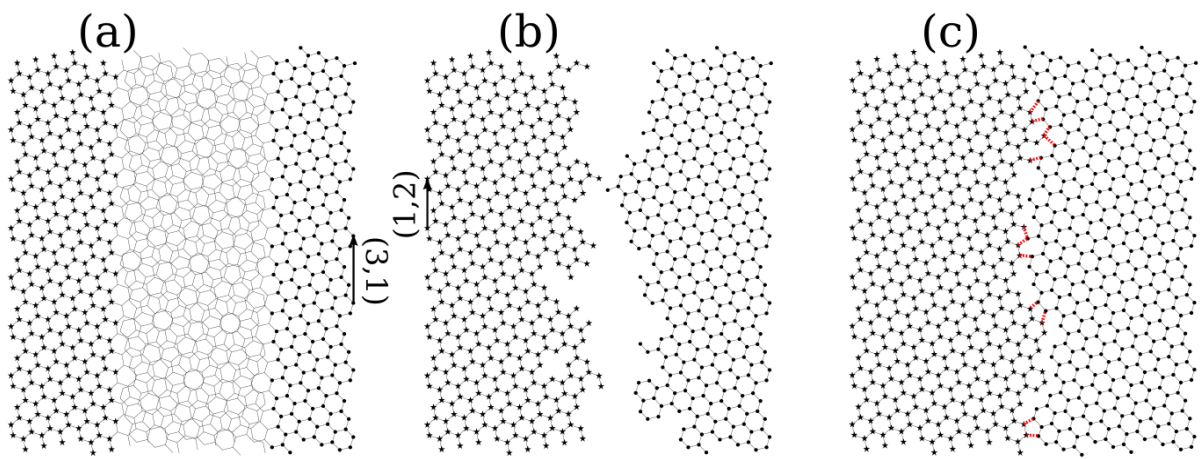

Fig. 2 - Construction of a disordered grain boundary between two graphene cells oriented such that their translation vectors (1,2) (left-hand side grain) and (3,1) (right-hand side grain) are parallel to one another. Occupied sites are represented by star symbols (left-hand side lattice) and filled circles (right-hand side lattice). The lattices of the cells overlap over a region where all the atoms have been removed (moiré pattern in panel (a)). In panel (b), 250 atoms have been added in the transition region, following the algorithm given in the text. In panel (c), the transition region has been filled in with 401 atoms. Possible bonds between the two grains, before any structural relaxation, are shown by thick dashed lines. The final structure is determined by geometry optimization.

Each time a site is filled with a C atom, a static random offset of its two coordinates is introduced, with small but increasing magnitudes as the distance to the geometrical border between the grains decreases. The reason for applying such a geometrical disorder is to fade away the moiré effect between the grain lattices in the interfacial region.

When an empty site lies close to C atoms of the other grain, its filling weight is reset to 0 whenever one of the following topological conditions is met: (a) the distance d to any atom is smaller than, $0.9 \cdot d_{cc}$ where $d_{cc} = 0.142$ nm is the C-C bond length in perfect graphene (to prevent unrealistically short CC distance across the border); (b) the number of C atoms in the first coordination shell ($0.9 \cdot d_{cc} \leq d < 1.35 \cdot d_{cc}$) is greater than 3 (to avoid atoms with more than three nearest neighbors); (c) the number of C atoms in the second coordination shell ($1.35 \cdot d_{cc} \leq d < 1.85 \cdot d_{cc}$) is greater than 6 (to avoid more than 6 next-nearest neighbors). These conditions prevent the growing grains to interpenetrate.

The filling procedure is continued until all the remaining empty sites have received a zero filling weight (Fig. 2c). Bonds are then formed to interconnect the two grains. Inter-grain bonds are allowed between atoms of both grains that have less than three nearest neighbors in their own grain lattice. The energy of the allowed inter-grain bonds is evaluated with a Keating-like potential[28]. Those bonds that lie below a given energy threshold are kept. Possibly, additional C atoms are placed at the center of triangles, if any, defined by twofold coordinated atoms and having approximately an equilateral shape with edge $\sqrt{3} \cdot d_{cc}$. Vacancies in the GB region are thereby avoided.

The full procedure is repeated several times, from the very beginning, for different, randomly chosen relative positions of the starting grains. The structure that is eventually selected is the one that has led to the smallest mean square deviation of the inter-grain bond lengths with respect to $d_{cc}$. The total energy of the selected structure is finally relaxed using the Tersoff-Brenner potential while keeping the structure flat, assuming it is bound to a substrate surface.

2.2 Wave Packet Dynamics

In our transport calculations the dynamics of the electrons while passing through the GBs were calculated by the WPD method. The time development of the $\psi(\vec{r},t)$ wave function is computed from the time dependent 3D Schrödinger equation using the split operator Fourier-transform method[29]

$$\psi(\vec{r},t+\Delta t) = e^{-i\hat{H}\Delta t}\psi(\vec{r},t) \tag{1}$$

$$e^{-i(\hat{K}+\hat{V})\Delta t} = e^{-i\hat{K}\Delta t/2}e^{-i\hat{V}\Delta t}e^{-i\hat{K}\Delta t/2} + O(\Delta t^3) \tag{2}$$

where the potential energy propagator is a simple multiplication with $\exp(-iV(\vec{r})\Delta t)$ for local potentials, and the effect of the kinetic energy propagator $\exp(-i\hat{K}\Delta t/2)$ is given in $k$ space by multiplicating the momentum space wave function $\phi(\vec{k},t)$ by $\exp(-i|\vec{k}|^2\Delta t/4)$. The input parameters of the WPD method are the potential $V(\vec{r})$ of the system and the $\psi(\vec{r},t_0)$ initial wave packet. From the calculated $\psi(\vec{r},t)$ wave function (output) we are able to obtain all measurable quantities, such as the probability density $\rho(\vec{r},t)$, the probability current density $j(\vec{r},t)$, etc. One of the advantages of the split operator method is the norm conservation of the wave packet, which is necessary to calculate accurate transport values during the simulation time.

Our potential model $V(\vec{r})$ consists of two parts: a metallic STM tip and the graphene surface containing GBs (Fig. 1). The metallic STM tip is approximated by a jellium potential, see Ref.

[23] for its parameters. For graphene, we used a local one-electron pseudopotential[30] matching the band structure of the graphene sheet π electrons. The π electron approximation is valid as long as the structure remains flat, as it was observed for graphene with grain boundaries growing on a flat substrate by AFM measurements[6,8]. The small deviation from the plane at the GB region was neglected in first approximation, precisely to preserve the validity of the pseudo-potentials.

Electrons represented by Gaussian wave packets were launched from the tip bulk towards the apex of the tip with momentum equal to the Fermi momentum: $\vec{k} = (0,0,-k_F)$. The real space width of the wave packet was $\Delta x, y, z = 0.37$ nm. This value meets two important criteria: i) its energy spread contains the region of interest of the graphene π band (±3 eV) and ii) $\Delta x, y, z = 0.37$ nm is significantly larger than the width of the STM-tip–graphene-tunneling channel, which was $\Delta x, y = 0.108$ nm in our calculation. In order to study the dynamics in the energy domain from only one time development calculation, we used a time-energy (t→E) Fourier transform $\psi(\vec{r},E) = F[\psi(\vec{r},t)]$. The energy resolution can be arbitrarily increased by increasing the total simulated time. The value of the energy resolution was set to 0.1eV as a good compromise between accuracy, runtime, and memory requirement.

2.3 Electronic structure calculations

Density of states (DOS) functions of ordered GBs were studied with the tight-binding (TB) and ab-initio density functional theory (DFT) methods. We used the π tight-binding Hamiltonian with first-neighbor hopping interactions $V_{ij} = -\gamma_0 (\frac{d_{CC}}{d_{ij}})^2$ where $\gamma_0 = +2.7 eV$ and $d_{ij}$ is the distance between the atoms i and j. The LDOS have been computed for the flat structure with the recursion method[31]. In order to investigate the effect of the deviation from the flat structure and the applicability of the π electron approximation we also performed DFT calculations within the framework of local density approximation (LDA) with the VASP simulation package[32,33]. Atomic positions were relaxed using the conjugate–gradient method in all dimensions. Projector augmented wave (PAW) pseudo-potentials[34,35] were used and the kinetic cut-off for the plane waves expansion was 400eV. The Brillouin-zone was sampled with a 2x12x1 and 4x14x1 Γ-centered k-space grid for the 5-7 and the 5-5-8 GB case, respectively. Two parallel 5-7 GBs were put in a rectangular super-cell to achieve periodical boundary conditions (see also Ref. [11]). Comparing the results of the two methods our assumption is confirmed about the applicability of the π electron approximation.

### 3. Results and discussion

#### 3.1 Ordered grain boundaries – the 5-7 grain boundary

The first analyzed ordered structure was an experimentally observed[14] periodical pentagon-heptagon (5-7) GB. In the simulation the STM tip was situated at 0.24 nm above the graphene surface and 2.5 nm away from the GB line on the right-hand side (Fig. 1). In Fig. 3 two snapshots can be seen selected from the time evolution of the probability density $\rho(\vec{r},t)$. These two particular time instants were chosen as moments when the WP has already tunneled from the tip apex into the graphene (t = 2.5 fs) and when the WP has already reached the GB line and started to pass through it (t = 5.2 fs). Fig. 3a shows that during the initial tunneling period between the STM tip and the graphene surface the tunneled WP on the graphene is spreading following the directions of the nearest neighbor C-C bonds. The lattice symmetry causes a six-fold shaped anisotropy on the spreading pattern.

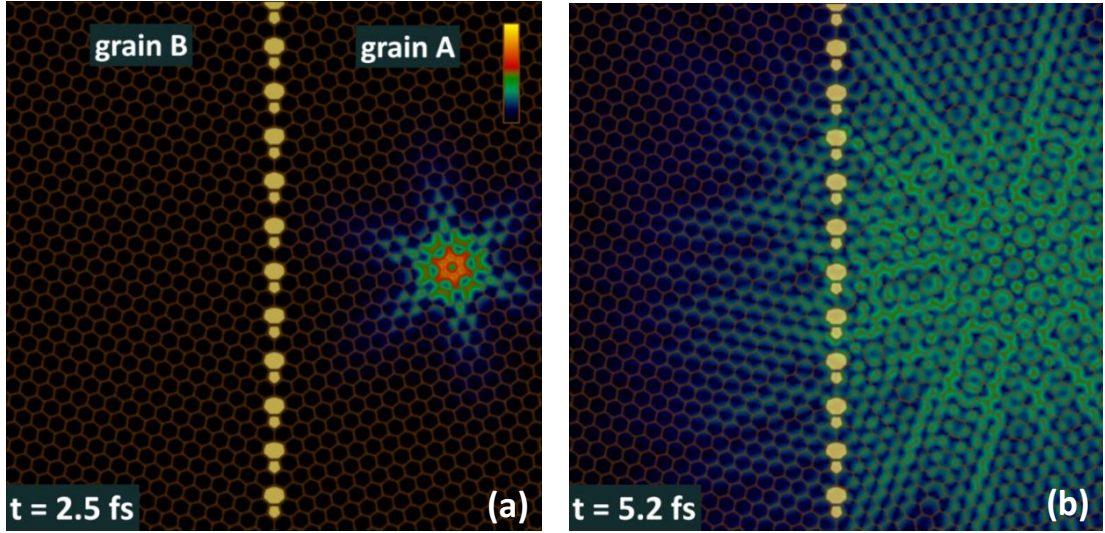

Fig. 3 - Selected snapshots from the time evolution of the probability density of wave packet shown as color-coded 2D (XY) sections. The STM tip is above the center of a hexagon in the right-hand side grain 2.5 nm away from the GB line. The graphene network is shown by thin orange lines, the non-hexagonal rings are highlighted. We used a nonlinear colour scale [see the scale bar in (a)] and different renormalization for (a) and (b) images. Black corresponds to zero, yellow to the maximum density. The size of the presentation window is 7.68 nm.

The calculation of the transmission coefficient for the GB is illustrated on Fig. 4. After the WP is injected into the right side grain of the graphene sheet, it spreads along the graphene sheet (see Fig. 3a). When the WP hits the GB, part of it is reflected back into the right side grain, but part of it is transmitted through the barrier into the left side grain (Fig. 4a). The transmission coefficient of the barrier is defined as $T(E) = I_{trans}(E)/I_{inc}(E)$, where $I(E)$ denotes the energy dependent probability current. The numerically calculated 3D wave function of the right hand side grain, however, contains an interference of the incident and reflected wave functions, as clearly seen on Fig. 3b and Fig. 8a. We can separate the incident wave function from this mixed state by repeating the calculation for a reference system containing no barrier, just a graphene sheet and an STM tip, as illustrated on Fig. 4b.

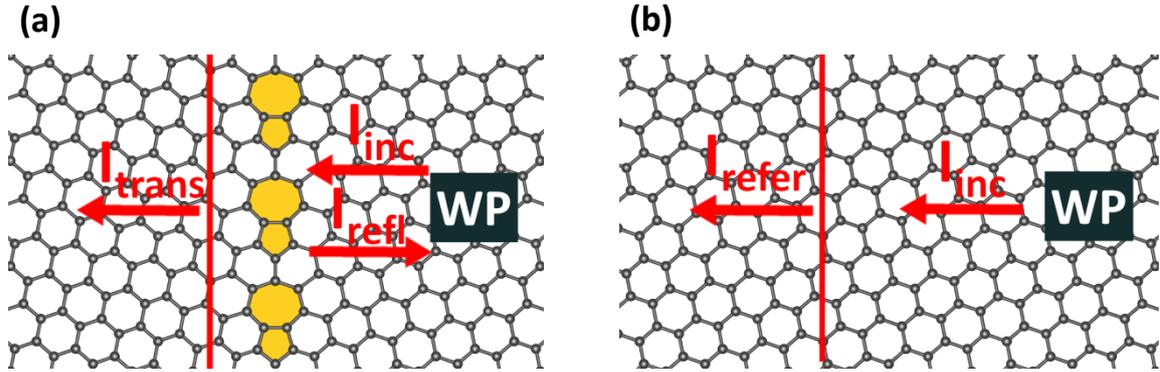

Fig. 4 – Schematic illustration of the transport calculation. (a) Geometry of the 5-7 GB where red arrows indicate the incident, reflected and transmitted probability currents. Red line shows the selected measurement plane for the transmitted probability current density. (b) Geometry of the reference system, which contains no GB. See the text for details.

Due to the same graphene lattice and same position of the STM tip $I_{inc}(E)$ is the same for the two systems, but the probability current in the reference system suffers no scattering, hence there are no transmitted and reflected currents, $I_{inc}(E) = I_{refer}(E)$. Thus the transmission coefficient was calculated as $T(E) = I_{trans}(E) / I_{refer}(E)$, where $I_{trans}(E)$ is the probability current calculated for the "real" system (that containing the GB), crossing the measurement plane at the left side of the GB (denoted by a red line on Fig. 4a) and $I_{refer}(E)$ is the probability current calculated for the "reference" system (that without the GB), where the current measurement plane in the "reference" system (denoted by a red line on Fig. 4b) is situated at the same place as for the "real" system.

We performed transport calculations for two different STM tip positions: i) over a C atom and ii) over the center of a hexagon, because the three-fold and six-fold symmetry of these STM tip positions have a major influence on the delocalized part of the wave functions in the far-field region[23](Fig. 1). The graphene clusters below the STM tip (near-field region) work as an energy filter for the incoming WP determining the further spreading in the far-field region. The vanishing LDOS at the Fermi energy in the case of the initial six-fold symmetry clusters[36] decreases the probability current around one order of magnitude in the far-field region as compared to the case of the tip situated over an atom.

The calculated transmission functions of the 5-7 GB for the two STM tip positions are shown in Fig. 5a. The result indicates that for those energy values where the GB structure has high DOS values ±0.4eV, ±1eV (Fig. 5b), the transmission is increased for both STM tip positions. These higher transmission values can be explained by an increased number of conductance channels according to the Landauer transport theory. The different transmission values around the Fermi level for the two STM tip positions are derived from the different initial symmetry of the two systems leading to differences in the probability current in the far-field region mentioned above.

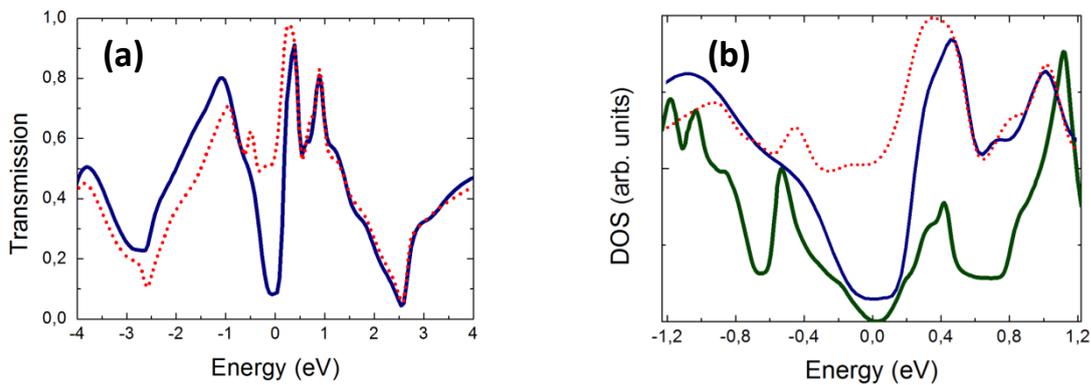

Fig. 5 - (a) Transmission function of the 5-7 GB calculated by WPD method. Red dotted (blue solid) line corresponds to the geometry when the STM tip is above an atom (a center of a hexagon), respectively. (b) *Ab-initio* DOS function (green solid line) of the 5-7 GB

superstructure and the transmission functions (Fig. 4a) in the vicinity of the Fermi energy. For those energies where the GB has DOS peaks the transmission functions have increased values.

A geometrical effect was observed as well, based on the mismatch angles of the two grains, which reduces the transmission in the low and high energy regions. At these „hot" energy levels ($\pm\gamma_0$, which correspond to Van Hove singularities in the perfect graphene π DOS, where $\gamma_0$ is the tight-binding interaction integral, see Sec. 2.3) the trigonal warping effect becomes enhanced due to the hexagonal distortion of the isoenergy curves of the graphene dispersion relation causing an anisotropic spreading of electrons along the zig-zag directions[23]. Fig. 6 shows the anisotropic spreading of the WP on both sides of the GB at $E = -\gamma_0$. The reduced transmission can be explained by taking into account the different orientation angles which implies different zig-zag directions of the two grains. The transition between the different zig-zag directions is realized by beam splitting at the GB. The geometrical effect is independent from the detailed atomic structure of the GB and relates only to the lattice mismatch of the two graphene grains. Furthermore, the increased reflection value is able to confine electrons between two parallel GBs, thus this system can work as an electronic waveguide[37].

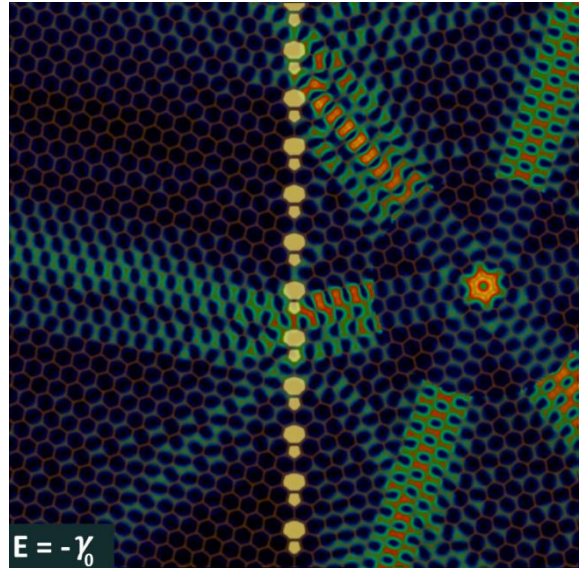

Fig. 6 - Probability density on the graphene sheet with the 5-7 GB for $E = -\gamma_0$ as a colour coded 2D (XY) section. Different colour scales were used in the near- and far regions (inside and outside of the circle). The GB works as a beam-splitter for the electrons spreading anisotropically along the zig-zag directions.

3.2 Ordered grain boundaries – the 5-5-8 line defect

As we mentioned, transmission properties across the GB depend on the electronic structure and the misorientation angle of the two grains. In the previous system (5-7 GB) we were able to study these effects simultaneously. In order to separate transport components we investigated an extended linear defect containing pentagons and octagons (5-5-8 defect line) where the mismatch angle is zero (Fig. 7a). This structure has been observed[13] previously in graphene grown on Ni surface and was shown to have a metallic behavior corresponding to the zigzag edges of the two sides. Due to the absence of the geometrical effect, we turned our attention to the interesting electronic structure of the 5-5-8 defect line. Tight-binding LDOS calculations of the system revealed that only atoms belonging to one of the graphene sublattices have finite LDOS value at the Fermi level (Fig. 7c). The atoms situated in the zigzag edge of the two grains (Fig. 7a) determine what we call the "metallic sublattice".

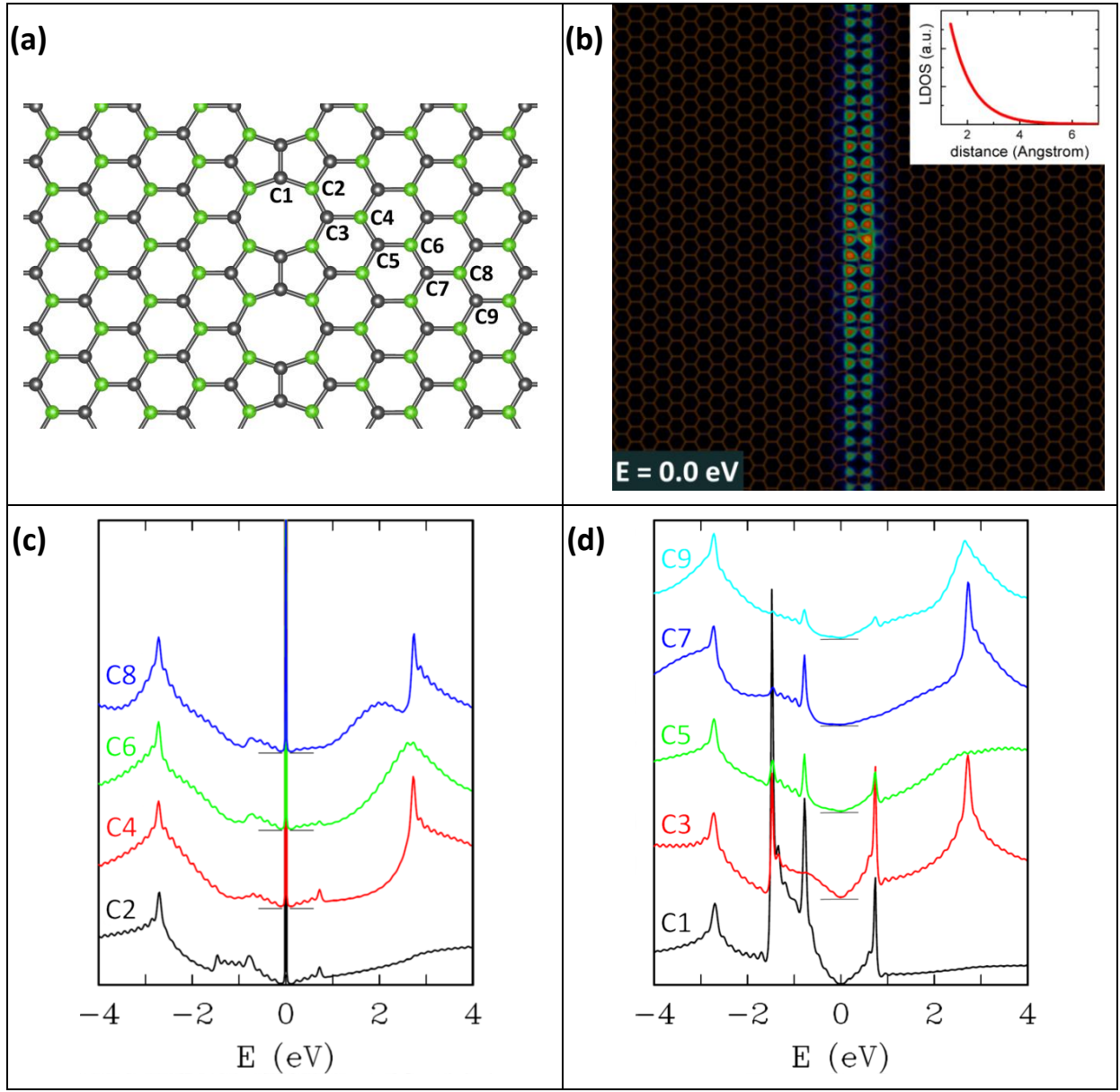

Fig. 7 – (a) Model geometry of the 5-5-8 line defect. Atoms marked in green belong to the sublattice having finite LDOS values at the Fermi energy. (b) Metallic wire behavior of the 5-5-8 defect line at the Fermi energy. Probability density on the graphene sheet shown as a colour coded 2D (XY) section. STM tip is above a C atom which belongs to the colored sublattice. WP is spreading along the line defect and only the metallic sublattice atoms

(marked by green) appear in the probability density image. Inset shows the exponential decay of LDOS values away from the line defect at the Fermi energy. (c)-(d) TB LDOS values of the atoms indicated by C1-C9 belonging to the two different sublattices.

For this system we analyzed two situations: i) when the STM tip is situated above the line defect, thus the WP spreads in a narrow channel along the defect and ii) when the STM tip is situated far from the defect, thus the WP crosses the defect. The atoms on the metallic sublattice (marked by green in Fig. 7a) are involved in the conduction at the Fermi energy in both cases.

In the first geometry our WPD simulations reveal that the WP has a strong localization around the defect line, with an exponential decay away from it, like in a waveguide (Fig. 7b). The width of the transport channel is determined by the finite LDOS values at the Fermi level near the defect. Indeed, TB calculations show that the LDOS functions have an exponential decay away from the defect line (Fig. 7b inset). Therefore, the metallic behavior of the system disappears at about 0.4 nm from the geometric center of the defect, opening a transport channel with approximately 0.8 nm width. The calculated width of the narrow channel by TB and WPD methods is in agreement with the decay length of the local DOS measured by STM[13]. It is worth reporting here that a recent calculation[38] found that if the line defect is doped, the atoms having LDOS value at the Fermi energy have also concentrated spin densities which induce a ferromagnetic state. The magnetic state can also influence the transport properties of the 5-5-8 defect line near the Fermi level.

In order to study the transport across the 5-5-8 defect line we placed the STM tip above the right-hand side grain, far from the defect. Despite of the zero misorientation angle of the two grains, the transmission value (Fig. 8a) is decreased above the Fermi energy at $E = +\gamma_0$ as observed in the 5-7 GB, but we see no such decrease around $E = -\gamma_0$. This transmission asymmetry for low and high energies is displayed in Fig. 8b and 8c.

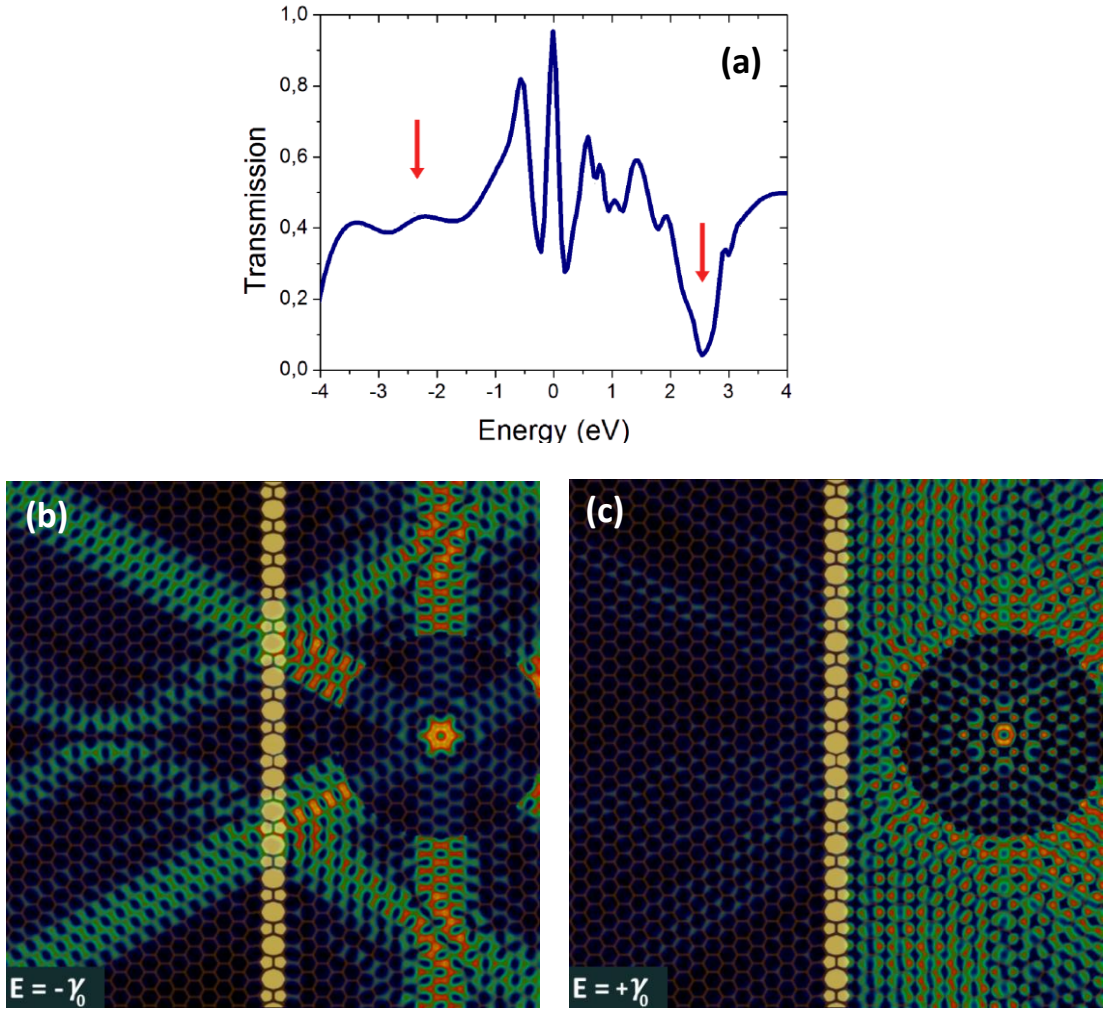

Fig. 8 – (a) Transmission function for the 5-5-8 defect line. The STM tip is above the center of a hexagon. Suppressed transmission appears above the Fermi energy at $E = +\gamma_0$. (b)-(c) Probability density on the graphene sheet for low and high energy regions (indicated with red arrows in (a)) shown as color coded 2D (XY) sections. Reduced LDOS values (Fig. 7) on the octagon atoms at $E = +\gamma_0$ suppress the probability denisty on the defect and decrease the transmission values across the 5-5-8 line defect. The images (b) and (c) are renormalized individually to their maximum density. The far region is shown on a separate (enhanced) colour scale.

The origin of the decreased transmission is not related to the geometrical effect like in the 5-7 GB, because as it can be seen in Fig. 8c the probability density of the WP does not enter into the extended line defect, but it reflects before reaching the left-hand side grain. The reflection of the WP causes a parallel current running along the right-hand side of the line defect. Therefore, we supposed that the asymmetry of the transport originates from the electronic structure of the 5-5-8 line defect. LDOS calculations proved our assumption and indicated an electron-hole asymmetry in the studied energy values. Fig. 7c and 7d show that the atoms of the octagon marked with 'C1' and 'C2' do not have LDOS peak at $E = +\gamma_0$ compared to the other atoms. However, the same peak occurs at $E = -\gamma_0$ on each atom. This electron-hole asymmetry appears only on the 'C1' and 'C2' octagon atoms where the reduced LDOS at $E = +\gamma_0$ causes the reflection (Fig. 8c) and the transmission minimum (Fig. 8a). All of our TB LDOS results in the 5-5-8 defect line have been verified with ab initio DFT calculations.

### 3.3 Disordered grain boundaries

It has been reported[9] that CVD graphene samples contain amorphous like grain boundaries and structural models[18,39] also predict different type of polygonal rings besides hexagons in the structure without periodicity. We modeled the disordered geometry along the way described in Sec. 2.1. In order to investigate the effect of disordered atomic structure of the GB on the transport properties we chose a large mismatch angle similar to the case of 5-7 GB. Fig. 9a displays the simulated disordered GB geometry. This large angle GB contains non-hexagonal rings from squares to nonagons, and the distribution of the polygons is similar to the one calculated by Malola et al [17]. Transport through the disordered GB has strong modulation around the Fermi energy. Fig. 9b shows transmission values compared to the ordered 5-7 GB.

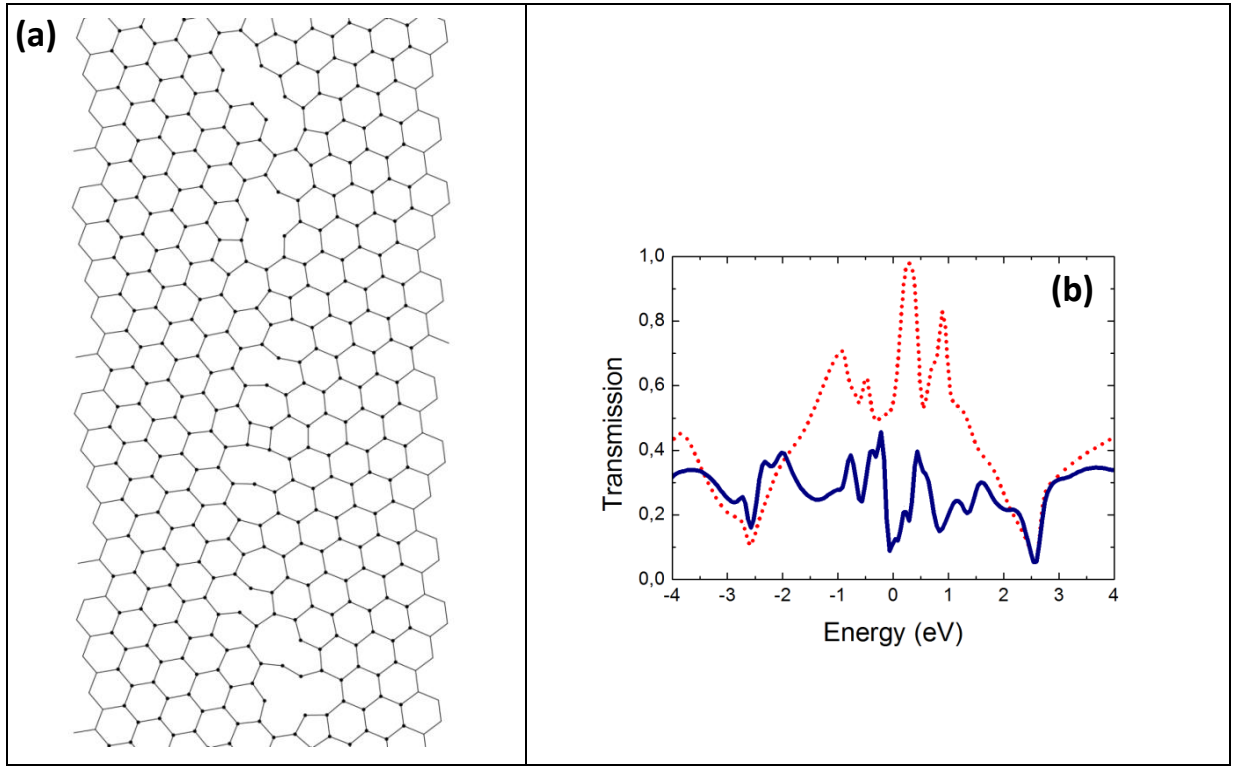

Fig. 9 – (a) Model geometry of the disordered GB. (b) Transmission functions of the ordered and disordered GB. Red dotted (blue solid) line corresponds to the ordered 5-7 GB (disordered GB) geometry. Reduced transmissions appear in the case of the disordered GB from -2eV to +2eV compared to the ordered one.

The same values at $E = -\gamma_0$ and $E = +\gamma_0$ verify our previous observation that the transmission depends solely on the misorientation angle of the grains in the energy domain where the electrons are spreading anisotropically along the zig-zag directions. From -2eV to +2eV we observed a significant decrease of the transmission values compared to the ordered case. The simultaneous presence of different polygons and two coordinated atoms alters dramatically the electronic structure of the GB and are sources of scattering centers leading to a higher resistivity. Near the Fermi energy the transmission decreases by about 75% which is in good agreement with the experimental measurements on individual GBs.[8]

The transport values depending on the concentration of the non-hexagonal rings have been also compared in the examined ordered and disordered GBs. A 11.5 nm$^2$ rectangular area was taken into account around the GBs, where the concentrations values were 9.7%, 18.3%, 19.4% for the ordered 5-7, 5-5-8 and disordered GBs, respectively. The corresponding transport values were 1.35, 1.12, and 0.45 as a result of the integration of the transmission functions $\int T(E)dE$ (dE=0.1 eV) for the low-energy carriers (EF ±1 eV). Thus, an increased concentration of non-hexagonal rings leads to reduced transport values. However, despite of the similar concentration for the 5-5-8 and disordered GB, the significantly different transport values emphasize the importance of the two coordinated atoms and different non-hexagonal polygons.

The influence of different stitching between grains on the electronic transport has been reported in recent experimental measurements[40]. Investigation of the degree of disorder is outside the scope of our paper, nevertheless the significance of disordered GBs on the transport properties is demonstrated by comparing ordered and disordered GBs.

## 4 Conclusions

We performed wave packet dynamical transport calculations for ordered and disordered graphene GBs. Our results for ordered periodic GBs highlight that the transmission properties are sensitive not only to the orientation of the adjacent grains but also to the detailed electronic structure of the GB. We also investigated non-periodical, disordered GBs which are more likely to be present in the CVD produced graphene samples. We developed a method to model the geometry of such structures. Transport simulations allow us to compare the electron transmission of one of such disordered structures to the ordered one. The suppressed transmission values for low-energy carriers explain the recent transport measurements across GB in CVD samples[8]. Further examination of different disordered GBs and the "defect engineering" of their structure by controlling the parameters of the CVD process may open a new way for further developments in high mobility graphene-based nanoelectronic devices.

**Acknowledgment**

The work in Hungary was supported in part by OTKA grant K101599 and an EU Marie Curie International Research Staff Exchange Scheme Fellowship within the 7th European Community Framework Programme (MC-IRSES proposal 318617 FAEMCAR project), the collaboration of Korean and Hungarian scientists was supported by the KRCF in the framework of the Korean-Hungarian Joint Laboratory for Nanosciences, the collaboration of Belgian and Hungarian scientists was supported by the bilateral agreement of the FNRS and HAS.